\documentclass[conference]{IEEEtran}
\IEEEoverridecommandlockouts

\usepackage{cite}
\usepackage{amsmath,amssymb,amsfonts}
\usepackage{algorithm}
\usepackage{algpseudocode}
\usepackage{textcomp}
\usepackage[utf8]{inputenc}
\usepackage[english]{babel}
\usepackage[pdftex]{graphicx}
\usepackage{pgfplots}
\pgfplotsset{compat=1.17}
\usepackage{pgfplotstable}
\usepackage{filecontents}
\usepackage{subcaption}

\newcommand{\lFig}[1]{\label{fig:#1}}
\newcommand{\lAlg}[1]{\label{alg:#1}}
\newcommand{\rFig}[1]{Fig. \ref{fig:#1}}
\newcommand{\lSec}[1]{\label{sec:#1}}

\newcommand{\rAlg}[1]{Algorithm \ref{alg:#1}}

\usepackage{xcolor}
\def\BibTeX{{\rm B\kern-.05em{\sc i\kern-.025em b}\kern-.08em
    T\kern-.1667em\lower.7ex\hbox{E}\kern-.125emX}}
    
\usepackage{eso-pic}
\newcommand\AtPageUpperMyright[1]{\AtPageUpperLeft{%
 \put(\LenToUnit{0.5\paperwidth},\LenToUnit{-1cm}){%
     \parbox{0.5\textwidth}{\raggedleft\fontsize{9}{11}\selectfont #1}}%
 }}%
\newcommand{\conf}[1]{%
\AddToShipoutPictureBG*{%
\AtPageUpperMyright{#1}
}
}

\begin{document}


\title{AppSlice: A system for application-centric design of 5G and edge computing applications
{\footnotesize \textsuperscript{}}
}

\author{\IEEEauthorblockN{Murugan Sankaradas}
\IEEEauthorblockA{
\textit{NEC Laboratories America}\\
Princeton, NJ \\
murugs@nec-labs.com}
\and
\IEEEauthorblockN{Kunal Rao}
\IEEEauthorblockA{
\textit{NEC Laboratories America}\\
Princeton, NJ \\
kunal@nec-labs.com}
\and
\IEEEauthorblockN{Srimat Chakradhar}
\IEEEauthorblockA{
\textit{NEC Laboratories America}\\
Princeton, NJ \\
chak@nec-labs.com}}

\maketitle
\conf{IEEE 12\textsuperscript{th} International Conference on Network of the Future, October 06-08, 2021,
Coimbra, Portugal}

\begin{abstract}
Applications that use edge computing and 5G to improve response times consume both compute and network resources. 
However, 5G networks manage only network resources without considering the application's compute requirements, and container orchestration frameworks manage only compute resources without considering the application's network requirements. We observe that there is a complex coupling between an application's compute and network usage, which can be leveraged to improve application performance and resource utilization. We propose a new, declarative abstraction called {\it app slice} that jointly considers the application's compute and network requirements. This abstraction leverages container management systems to  manage edge computing resources, and 5G network stacks to manage network resources, while the joint consideration of coupling between compute and network usage is explicitly managed by a new runtime system, which delivers the declarative semantics of the app slice. The runtime system also jointly manages the edge compute and network resource usage automatically across different edge computing environments and 5G networks by using two adaptive algorithms.
We implement a complex, real-world, real-time monitoring application using the proposed app slice abstraction, and demonstrate on a private 5G/LTE testbed that the proposed runtime system significantly improves the application performance and resource usage when compared with the case where the coupling between the compute and network resource usage is ignored.

\end{abstract}


\section{Introduction}
\label{introduction}
\begin{figure}[tb]
    \centering
    \includegraphics[width=0.72\linewidth,scale=0.55]{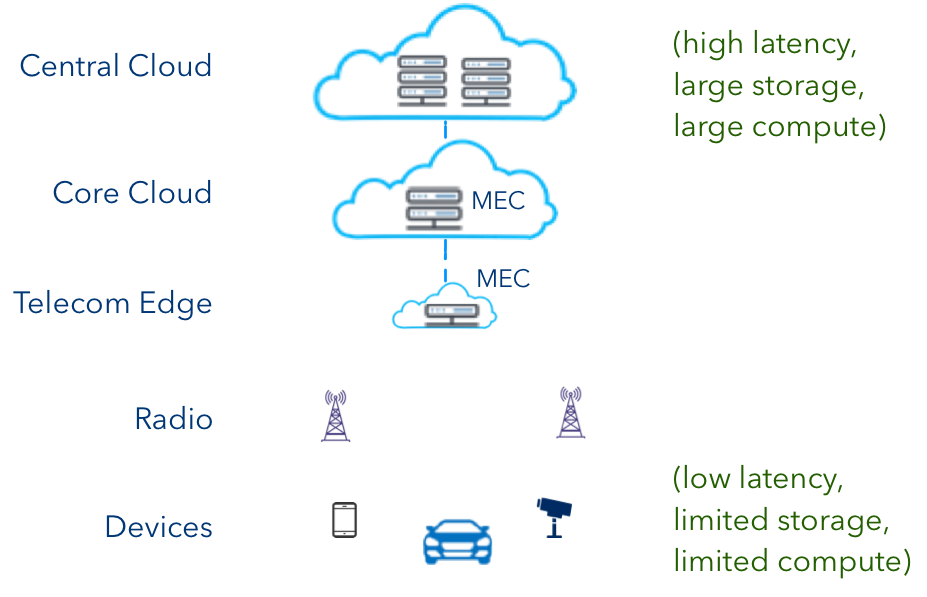}
    \caption{Compute and network tiers}
    \label{fig-network-compute-tiers}
\end{figure}

Edge computing and 5G are inextricably linked technologies that promise to enable huge amounts of data to be processed in real-time and to significantly improve the response times of low-latency applications. 
Edge computing \cite{cao_overview_2020} brings compute, storage, switching and control functions relatively close to end users and IoT endpoints. Emerging tiers of network and compute is shown in Figure~\ref{fig-network-compute-tiers}. In these tiers, critical data can be processed at the edge of the network, while less urgent data can be sent to the cloud for data processing. 


5G networks promise ultra-low latency, high reliability, high bandwidth and high device density, and they are expected to enable a wide variety of emerging applications like remote operation, remote maintenance, augmented reality, mobile workforce and enterprise applications (like payments, tactile, V2X \cite{9345798} and real-time surveillance). Often, these applications require low latency (0.5 to 10 milliseconds), high data rate (10 to 1000 Mbps), and high device density (1000s of sensing devices).




Although it is widely recognized that both edge computing and 5G networks are required to realize ultra low-latency applications, there is no principled, application-centric approach that automatically, and dynamically, optimizes the application across a range of edge computing platforms and 5G network stacks \cite{mec-netslice} \cite{aws-verizon-5g} \cite{edge-computing-in-5g}. Today, 5G networking vendors manage and provision network resources (soft or hard network slicing \cite{8845282}\cite{demystifying-network-slicing}) without considering the application's compute requirements \cite{9382385} \cite{9272281}. Similarly, container orchestration frameworks like Kubernetes \cite{kubernetes-link} manage and provision compute resources without considering the application's network requirements. It is also the case that these container management systems manage or provision compute resources only within a specific tier in the layered computing fabric shown in Figure \ref{fig-network-compute-tiers}, rather than across all tiers of the computing fabric.

In order to have a unified, application-centric view that jointly considers both compute and network resources for an application across diverse edge computing environments and 5G network stacks, we propose a new abstraction layer called {\it app slice}, which jointly considers the application's compute as well as the network resources. This abstraction layer leverages container management systems to  manage edge computing resources, and 5G network stacks to manage network resources, while the joint consideration of coupling between compute and network resource usage is explicitly managed by a new runtime for the proposed app slice abstraction.

We make the following key contributions in this paper.

\begin{itemize}
    \item We propose a new, application-centric declarative specification, called {\it app slice}, which allows joint specification of compute and network requirements of the application as a whole, as well as the individual functions (or microservices) that make up the application.
    \item We propose a new runtime system, which realizes the declarative semantics in the app slice specification, and jointly manages the edge compute and network resources across different edge computing environments and 5G networks by using two adaptive algorithms.
    \item We implement a complex, real-world, real-time monitoring application using the proposed app slice abstraction, and demonstrate on a private 5G/LTE testbed that the proposed app slice specification and runtime system significantly improve the application performance and resource usage when compared with the case where the complex coupling between the compute and network resource usage is ignored. 
\end{itemize}

\section{Declarative, App slice specification}
\lSec{slice-spec}


We model the application as a set of functions or microservices, specified through an {\it app specification}. App specification includes application details, the functions that comprise the entire application, function details along with the instances of functions and finally the inter-connection between various function instances. 

The proposed declarative {\it app slice specification} 
consists of two parts: an application-level specification that captures application-level requirements like latency and bandwidth, and a function-level specification that captures the compute and network requirements of each function.
    

\subsection{Application-level specification}
It captures the desired application requirements: 
\begin{itemize}
    \item \textit{latency}: End-to-end application latency (in milliseconds).
    \item \textit{bandwidth}: Overall network bandwidth (in Mbps).
    \item \textit{deviceCount}: Total number of devices the application connects to, or expects to receive data streams from.
    \item \textit{reliability}: Desired reliability (between 0 and 1). 0 being unreliable and 1 being totally reliable.
\end{itemize}

\subsection{Function-level specification}
It includes compute and network specification.
\subsubsection{Network specification}
The network requirements of a function are specified as part of the this specification.
\begin{itemize}
    \item \textit{latency}: Maximum tolerable network latency (in milliseconds). When function A’s output is fed to another function B as input, then latency specification for function B is the latency of the link connecting function A to function B. 
    \item \textit{throughputGBR}: Guaranteed network bandwidth (in Mbps) required by the function.
    \item \textit{throughputMBR}: Maximum bandwidth (in Mbps) that can be consumed by the function. 
    \item \textit{packetErrorRate}: Ratio of the number of incorrectly received packets and the total number of received packets.
    \item \textit{duration}: Optional duration (in milliseconds) for which the network guarantees should be provided for the function. Default value for this is ``auto", indicating that the runtime can choose and decide this value dynamically.
\end{itemize}

\subsubsection{Compute specification}
Function's compute requirements are specified as part of this specification.
\begin{itemize}
    \item \textit{minCPUCores}: Desired minimum CPU resources (absolute cpu units). 1 represents either 1 vCPU/core on the cloud or 1 hyperthread on bare-metal Intel processors. 1 cpu unit is divided into 1000 ``millicpus" and the finest granularity that can be specified is ``1m" (1 millicpu). 
    \item \textit{maxCPUCores}: Maximum CPU cores (absolute cpu units) that the function can use.
    \item \textit{minMemory}: Desired minimum memory (in bytes).
    \item \textit{maxMemory}: Maximum allowable memory (in bytes).
    \item \textit{tier}: Optional parameter to specify specific tier i.e. ``device", ``edge" or ``cloud" in the computing fabric where the function should run. Default value for this is ``auto", indicating that the function can run anywhere in the computing fabric.
\end{itemize}



\section{App slice runtime system (RS)}
\begin{figure}[b]
 \centering
    \includegraphics[height = 2.5 in]{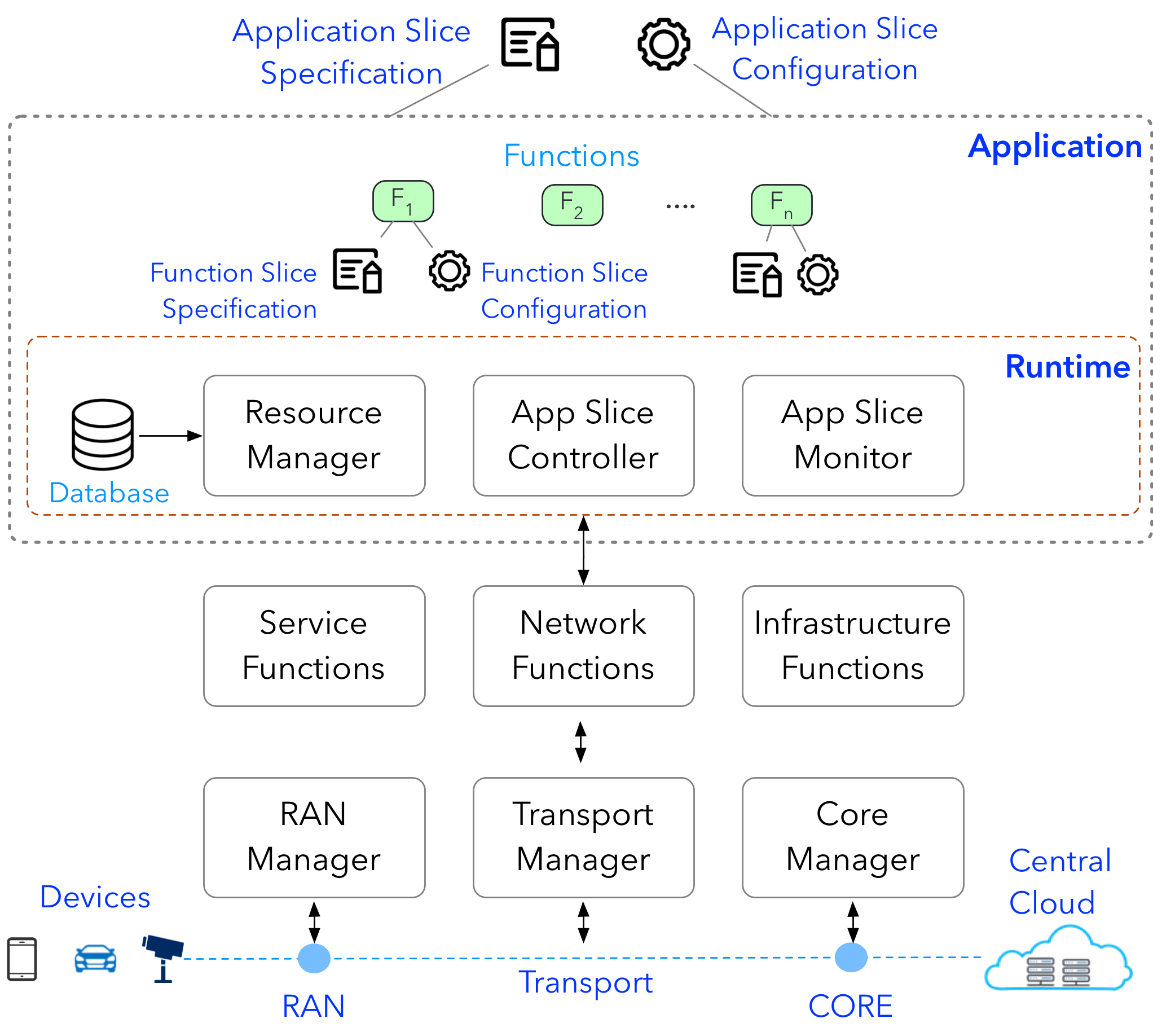}
    \caption{App slice runtime}
    \label{fig-runtime}
\end{figure}

\lSec{slice-runtime}
Our RS, shown in Figure \ref{fig-runtime}, (a) is integrated with the application itself (b) sits on top of the compute and 5G network infrastructure (c) relies on standard APIs from the underlying 5G network for network slices (d) relies on standard APIs from underlying compute infrastructure like Kubernetes \cite{kubernetes-link} for compute slices and (e) manages the overall execution of the application.
RS consists of Resource Manager, App Slice Controller and App Slice Monitor. 

\subsection{Resource Manager (RM)}
RM manages execution of total application by co-ordinating with ASC and ASM. 
\begin{algorithm}[t]
    \algnewcommand{\LineComment}[1]{\State \(\triangleright\) #1}
    \caption{RM resource allocation}
    \lAlg{rm-resource-allocation}
    \textbf{Input:} Resource request per function (for application)  \\
    \textbf{Output:} Allocated Resources per function (for application)
    \begin{algorithmic}[1]
    \LineComment{key: function, value: allocated tier resources}
    \State Initialize map $\rightarrow$ $resources$
    \For {$function$ $\in$ $functions$}
    \LineComment{Order tiers in ascending order of resource cost}
    \LineComment{Cheaper tiers are checked before expensive ones}
        \For {$tier$ $\in$ $tiers$}
        \LineComment{Match requested with available tier resources}
            \If {matchResoures($c\_r$, $n\_r$, $tc\_r$, $tn\_r$)}
                \LineComment{Allocate tier resources to function}
                \State $resources[app] \leftarrow$ $tc\_r$, $tn\_r$;
                \State break;
            \EndIf
        \EndFor
    \EndFor
    \LineComment{Return allocated resources}
    \State \Return $resources$
    \end{algorithmic}
\end{algorithm}
For resource allocation, RM first checks application level slice specifications. Then, for each function in the application, RM follows the algorithm shown in \rAlg{rm-resource-allocation}. First priority is given to meeting function's compute and network requirement and second priority is given to the cost. If the resource request cannot be met for the application and all its associated functions, then RM reports it to the application, and leaves it to the application and associated functions to take appropriate actions.

\begin{algorithm}[t]
    \algnewcommand{\LineComment}[1]{\State \(\triangleright\) #1}
    \caption{RM dynamic resource adjustment}
    \lAlg{rm-dynamic-resource-adjustment}
    \begin{algorithmic}[1]
    \While{true}
        \For {$function$ $\in$ $functions$}
            \LineComment{check if resource conditions have changed}
            \If {resourceConditionChanged($c\_r$, $n\_r$)}
                \LineComment{check if new resources are available}
                \State $resources \leftarrow$ getResources($c\_r$, $n\_r$)
                \If {$resources$}
                    \LineComment{schedule on new resources}
                    \State scheduleFunction($function$, $resources$)
                \Else
                    \LineComment{report error for the function}
                    \State reportError($function$)
                \EndIf
            \EndIf
        \EndFor
        \State sleep($interval$)
    \EndWhile
    \end{algorithmic}
\end{algorithm}

As various functions continue to run, RM periodically monitors the status of these functions and adjusts the resources, if needed. To do so, RM follows \rAlg{rm-dynamic-resource-adjustment}, where at every $interval$ seconds, which is configurable (in our experiments we set this interval to be 2 seconds), RM checks across all the functions that are executing. Specifically, RM checks if the resource requirements of a function are being met or not by the allocated tier's compute and network resources. If for whatever reason (change in operating conditions/input content, load burst, network disruption or hardware failure) the network or compute resources are found to be insufficient, then RM tries to find additional resources. 
Along with checking if additional resources are needed, RM also checks if resources are under-utilized, and if so, releases them. 


\subsection{App Slice Controller (ASC)}
\begin{figure}[t]
 \centering
    \includegraphics[width=0.8\linewidth]{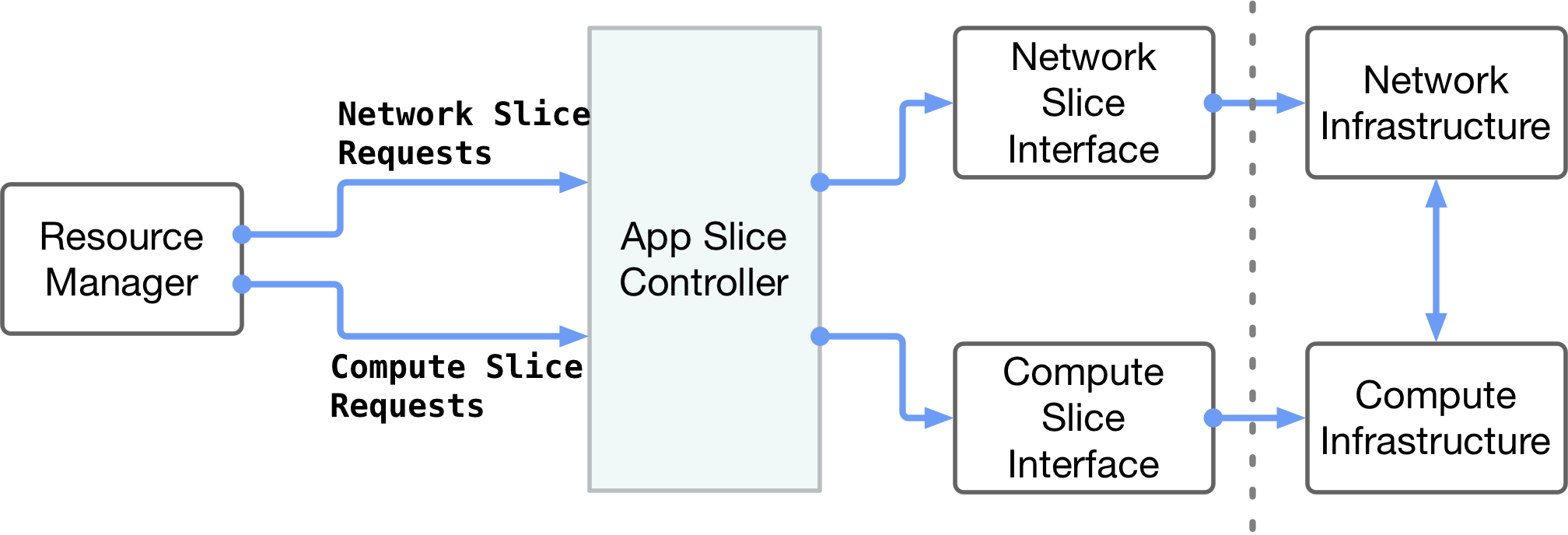}
    \caption{App slice Controller}
    \lFig{slice-controller}
\end{figure}
ASC, shown in \rFig{slice-controller}, follows directions from RM and manages the slicing, including compute and network slicing for functions. Although the name says ``controller", ASC does not actually ``control" low-level network or compute slices, rather it uses the underlying infrastructure to manage network and compute slices for various functions. For network slices, to enforce admission control, we create a custom layer on top of existing network vendors \cite{celona-link} and expose this layer to ASC. For compute slices, ASC directly interacts with the underlying compute infrastructure.

\subsection{App Slice Monitor (ASM)}
ASM continuously monitors and collects various compute and network usage metrics. Specifically, in our setup (described in Section \ref{testbed}), we have 5G network (and network slicing) between devices (CPE) and edge (MEC), and ASM uses the underlying 5G network’s APIs to measure latency, throughput and packet error rate. Beyond MEC and into the cloud, there is no 5G network (and therefore no network slicing) as it is over WAN. For this, ASM uses tools like iPerf3 \cite{iperf3} to collect network related metrics. With respect to collecting compute related metrics, ASM uses underlying compute infrastructure's APIs, in our setup Kubernetes' APIs. Note that unlike network slicing, compute slicing extends all the way to the cloud and there is Kubernetes cluster setup at each tier (devices, MEC and cloud). ASM communicates with appropriate Kubernetes cluster in the specific tier to collect compute related metrics.
These metrics are used by RM to manage application execution. 

\section{Experimental Setup}
\lSec{experiments}
\subsection{Testbed}
\label{testbed}
\begin{figure}[b]
    \centering
    \includegraphics[width=0.95\linewidth]{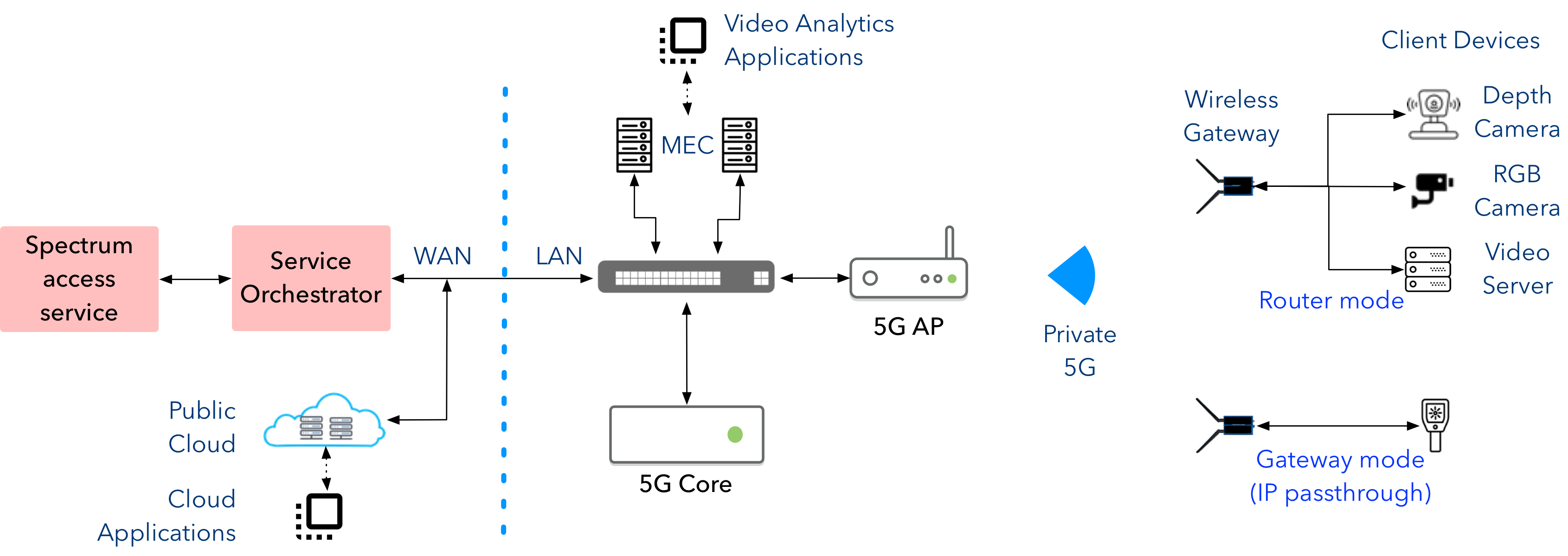}
    \caption{Architecture of testbed}
    \label{fig-testbed-architecture}
\end{figure}

A high level overview of our testbed is shown in Fig. \ref{fig-testbed-architecture}. 
In our testbed, we have used wireless gateways from Multitech \cite{multitech-link} to connect Customer Premise Equipment (CPE) (cameras and video servers) over private 5G to Access Points from Celona \cite{celona-link}. Data and control plane traffic from the access points is terminated at the Core (from Celona). Core is activated and configured remotely via Celona's Service Orchestrator. Organizations can set specific SLAs and network requirements for different device groups and application types using network slicing. We have Multi-access Edge Computing (MEC) \cite{7901477} servers connected in the same LAN with one master and three worker node servers. Master node is equipped with 10-core Intel core i9 CPU and the three worker nodes are equipped with 24-core Intel CPU and with NVIDIA RTX 2080 Ti GPUs.

\subsection{Application: Real-time monitoring (RTM)}
\label{real-time-monitoring}
We implemented and used RTM video analytics application for our experiments. RTM provides fast and reliable identification of pre-registered individuals using face recognition technology. Various functions of this application, along with the pipeline is shown in \rFig{realtime-monitoring-application}. ``Video Sensor" extracts frames, ``Face Detection" extracts faces, ``Feature Extraction" extracts unique facial features, ``Face Matching" provides face matches (using pre-registered face gallery from ``Biometrics Manager"), and finally these face matches are then stored and delivered as alerts through ``Alerts Manager". 


\begin{figure}[t]
    \centering
    \includegraphics[width=0.9\linewidth,scale=0.15]{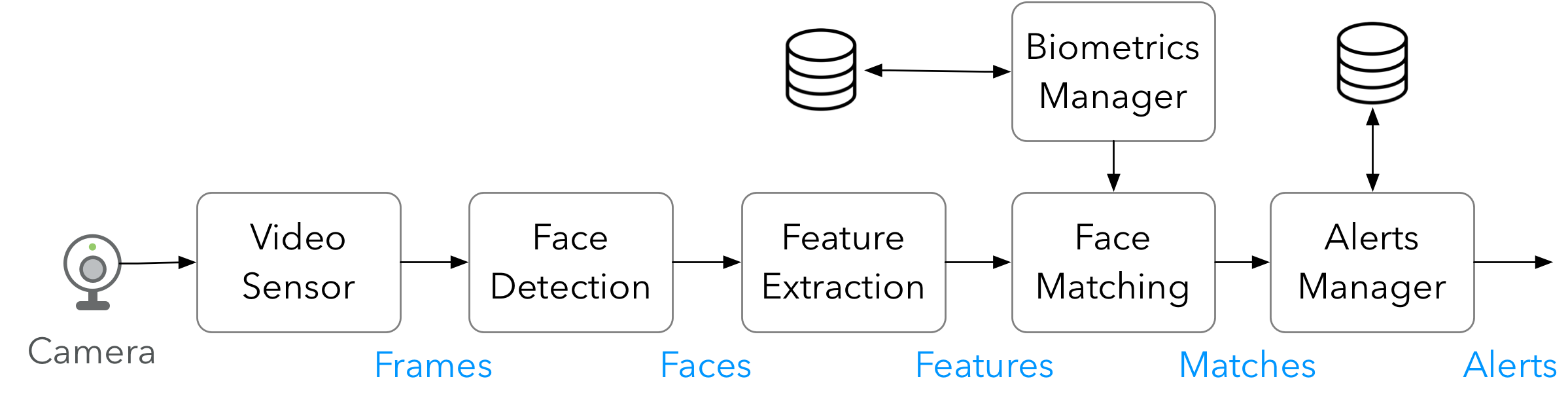}
    \vskip -0.1in
    \caption{Real-time monitoring}
    \lFig{realtime-monitoring-application}
\end{figure}

\subsection{Video streaming setup}
We used \textit{Labelled movie trailer dataset (LMTD)} \cite{10.1016/j.asoc.2017.08.029} for our experiments. We selected a video sequence containing 22 unique people from this dataset and used it as input video feed. 10 among the 22 were pre-registered in ``Biometrics manager". A combination of IP cameras (AXIS Q3515) and video streaming servers was used to generate large video traffic into our testbed. IP cameras were pointed to large display, which was looping video sequence.


\section{Results}
\lSec{results}
In order to study the impact of compute and network resources on the accuracy of insights from the application, we have to consider application-level metrics. Therefore, in our experiments, for RTM application, we chose to measure the total number of alerts produced by the application as a measure of performance of the application. Different applications can have different metrics.
\subsection{Performance without app slice}
\lSec{perf-without-app-slice}
\subsubsection{Impact of varying network}
\begin{figure}[tb]
\begin{subfigure}[]{0.5\linewidth}
\centering
    \includegraphics[height=0.85 in]{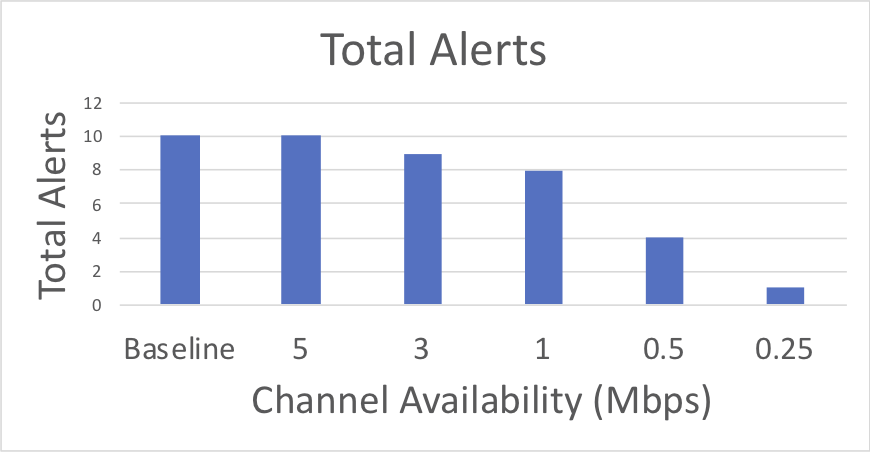}
    \caption{Impact of varying network}
    \label{impact-of-varying-network}
\end{subfigure}%
\begin{subfigure}[]{0.5\linewidth}
 \centering
    \includegraphics[height=0.85 in]{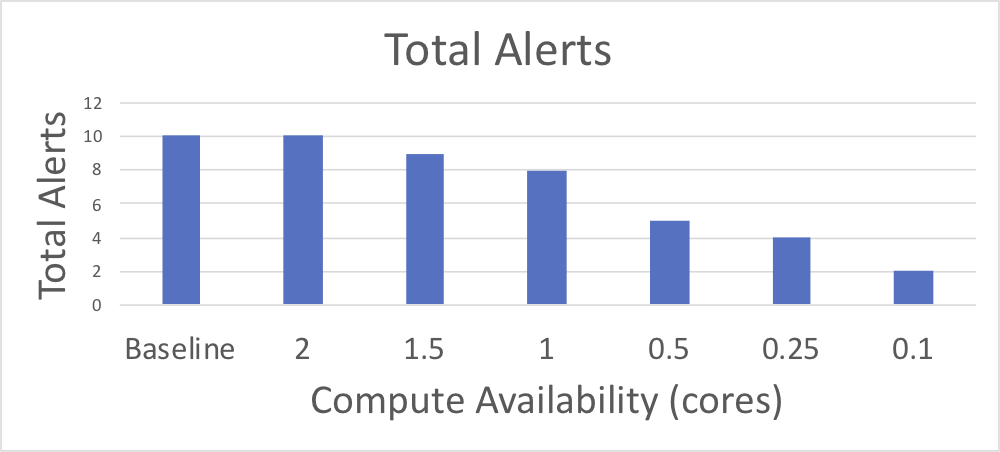}
    \caption{Impact of varying compute}
    \label{impact-of-varying-compute}
\end{subfigure}





\caption{Performance without app slice}
\label{perf-without-app-slice}
\end{figure}

Fig. \ref{impact-of-varying-network} shows the impact on application performance when the available network to the application varies from 5 Mbps to 0.25 Mbps (note that there is enough compute available for the application). To reduce the available network for the application, we manually pump additional traffic through the network, which impacts the application performance adversely. We see that the total number of alerts goes down from 10 to 1.


\subsubsection{Impact of varying compute}

Fig. \ref{impact-of-varying-compute} shows the impact on application performance as the available compute goes down from 2 to 0.1 (note that there is enough network available for the application). We did this for face detection component, which is compute-intensive function and critical in determining application-level accuracy. To reduce the available cores, we used stress-ng \cite{7877341}, which is a CPU load generation tool. We observe that as the available compute reduces, the total number of alerts drops from 10 all the way to 1. This is because fewer frames are processed as available compute goes down (frames with face match never get processed).


\subsubsection{Impact of varying compute and network}
\begin{figure}[tb]
\begin{subfigure}[]{0.5\linewidth}
\centering
    \includegraphics[height=0.85 in]{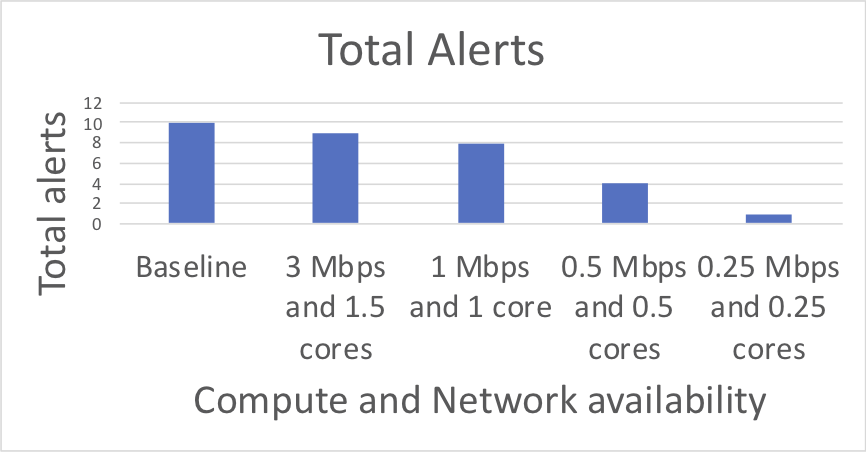}
    \caption{Network as bottleneck}
    \label{network-as-bottleneck}
\end{subfigure}%
\begin{subfigure}[]{0.5\linewidth}
    \centering
    \includegraphics[height=0.85 in]{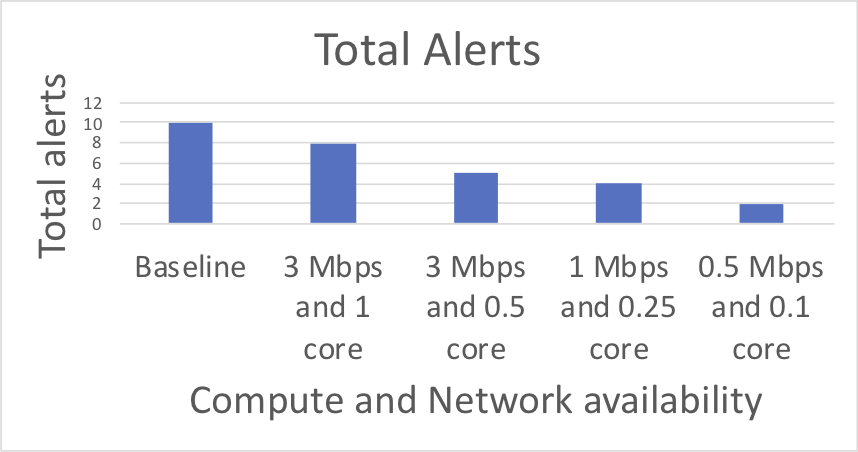}
    \caption{Compute as bottleneck}
    \label{compute-as-bottleneck}
\end{subfigure}
\caption{Impact of varying compute and network}
\label{vary-compute-and-network}
\end{figure}

Next, we varied the network as well as the compute resources at the same time and observed RTM application performance. Particularly, we observed the total number of alerts when we gradually changed the network resources down from 5 Mbps to 0.25 Mbps, and compute resources down from 2 cores to 0.1 cores in various combinations. We observed that as the network and compute resources went down, the total number of alerts also went down from 10 to 1 (degradation of 90 \%), as shown in Fig. \ref{vary-compute-and-network}. In Fig. \ref{network-as-bottleneck} we see that at 0.5 Mbps and 0.5 cores, although compute was capable of receiving 5 alerts (Fig. \ref{impact-of-varying-compute}), we only saw 4 alerts, indicating that network was bottleneck. Similarly, in Fig. \ref{compute-as-bottleneck}, we observe that at 3 Mbps and 1 cores, although network was capable of receiving 9 alerts (Fig. \ref{impact-of-varying-network}), we only saw 8 alerts, indicating that compute was bottleneck. Similar trend is observed for different combination where either network or compute can become bottleneck and have adverse effect on the application.




\subsection{Performance with app slice}
\lSec{perf-with-app-slice}
\begin{figure}[tb]
\begin{subfigure}[]{0.4\linewidth}
\centering
    \includegraphics[height=0.85 in]{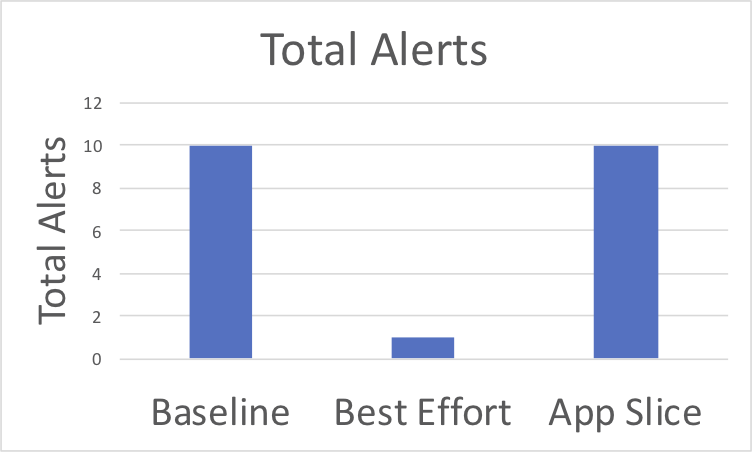}
    \caption{Impact of app slice}
    \label{fig-slicing-result}
\end{subfigure}%
\begin{subfigure}[]{0.6\linewidth}
\centering
    \includegraphics[height=0.85 in]{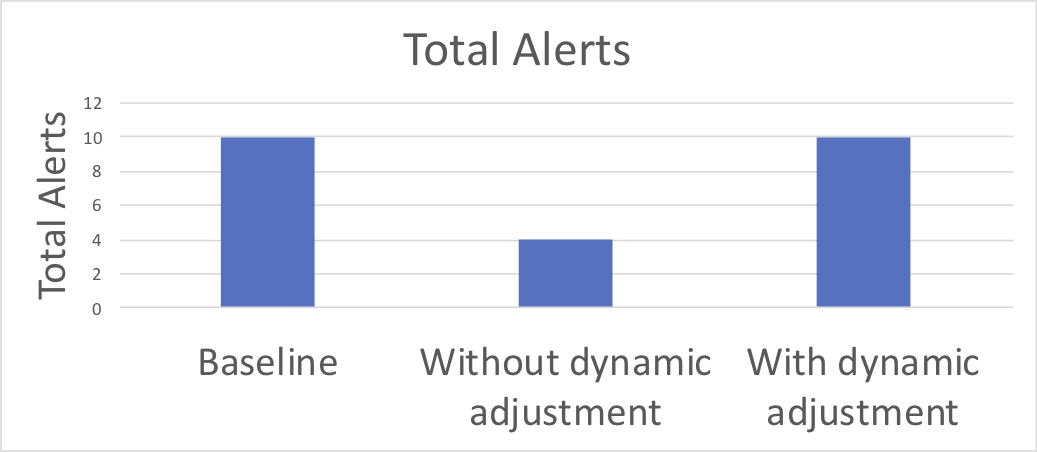}
    \caption{Dynamic resource adjustment}
    \label{fig-dynamic-result}
\end{subfigure}
\caption{Performance with app slice}
\label{perf-with-app-slice}
\end{figure}

We measured the performance of RTM application, once the app slice (which includes the network and compute requirements for each function) is specified and the application is run through our app slice RS. Note that app slice RS is placed at MEC in our setup. As before, we simulated a condition where there is external load and in such condition we conducted this experiment. As shown in Fig. \ref{fig-slicing-result}, we see that when there is no app slice specified i.e. under ``Best effort" conditions, the total number of alerts is only 1, whereas in the presence of ``App slice", the total number of alerts is 10 (same as baseline). 

Fig. \ref{fig-dynamic-result} shows the impact of dynamic resource adjustment. Here, we assume that based on the initial profiling of the application, the app-slice specified network bandwidth is 0.5 Mbps and compute is 2 cores. With this specification, and without any dynamic resource adjustment, we observed that only 4 alerts were received. However, when we use dynamic resource adjustment, our RS identifies (based on the compute and network resource usage) that network is the bottleneck and it increases the allocated network resources to 5 Mbps, while keeping the compute as 2 cores. This results in the number of received alerts increasing to 10 (which is same as the baseline). 

This shows that our RS understands the coupling between network and compute, and is able to dynamically adjust appropriate resources so that application performance is optimized. Our methodology and design are applicable to a wide range of video analytics applications and we use RTM application as an illustrative example.


\section{Related Work}
\lSec{related-work}

Exisiting standard specifications like TOSCA \cite{tosca} do apply to applications modelled as a collection of services. However, they are focused only on the cloud and do not extend to multiple tiers, like compute resources in devices, or edge (MEC). Network slicing applied to smart grid \cite{9389979} \cite{9152315} or healthcare \cite{10.1109/MNET.011.1900458} mainly relies only on network slice specification, which is about network and compute resources required to execute (virtualized or physical) network functions. However, applications on top of the network can use a variety of compute resources, which are not specified as part of the network slice specification. 

Another recent work \cite{li2016endtoend} also proposes to slice computation and communication resources. They propose vertical and horizontal slicing of the air interface, radio access network, core and the virtualized computing resources available to execute network function virtualization. However, there is no joint consideration of application's compute and network resource usage. To the best of our knowledge, our proposal is the first to jointly consider compute and network requirements to improve efficiency and performance of 5G applications across different edge compute environments and 5G network stacks. 


\section{Conclusion}
\lSec{conclusion}
Recognizing the complex coupling between compute and network usage of an application, we proposed a new, declarative abstraction called app slice. It allows joint consideration of compute and network requirements of 5G and edge computing applications. The declarative semantics of the new abstraction are realized by a new app slice runtime, which ensures that the application automatically optimizes the compute and network resource usage on different edge computing environments, and different private or carrier 5G networks. We also implemented a real-world,  real-time video analytics application using the app slice abstraction, and performed extensive experiments on a private 5G testbed to demonstrate the positive impact of the proposed abstraction and its runtime system on the performance and resource utilization of applications. 


\bibliographystyle{IEEEtran}

\end{document}